\documentclass{aa}

\usepackage{txfonts}
\usepackage{natbib}
\bibpunct{(}{)}{;}{a}{}{,} 
\usepackage{graphicx}

\usepackage{soul}

\begin{document}

\title{Variability in X-ray line ratios in helium-like ions 
of massive stars: the wind-driven case}

\titlerunning{Wind-driven X-ray line ratio variability}
\authorrunning{R. Ignace et al.}


\author{R. Ignace\inst{1}
\and
Z. Damrau\inst{1}
\and 
K. T. Hole\inst{2}
}

\institute{Department of Physics \& Astronomy, East Tennessee
State University, Johnson City, TN, 37614, USA
\and
Norwich University, 158 Harmon Drive, Northfield, VT, 05663, USA
}

\offprints{R. Ignace, \email{ignace@etsu.edu}}

\date{Received <date> / 
Accepted <date>}

 \abstract
   {
   }
   {
High spectral resolution and long exposure times are providing
unprecedented levels of data quality of massive stars at X-ray
wavelengths.  A key diagnostic of the X-ray emitting plasma
are the {\em fir} lines for He-like triplets.  In particular,
owing to radiative pumping effects, the forbidden-to-intercombination
line luminosity ratio, $R=f/i$, can be used to determine
the proximity of the hot plasma to the UV-bright photospheres
of massive stars.  Moreover, the era of large observing programs additionally
allows for investigation of line variability.
   }
   {
This contribution is the second to explore how variability in the
line ratio can provide new diagnostic information about distributed
X-rays in a massive star wind.  We focus on wind integration for
total line luminosities, taking account of radiative pumping and stellar
occultation.  While the case of a variable stellar radiation field
was explored in the first paper, here the effects of wind variability
are emphasized.
   }
   {
We formulate an expression for the ratio of line luminosities $f/i$ that
closely resembles the classic expression for the on-the-spot
result.  While there are many ways to drive variability
in the line ratio, we use variable mass loss as an illustrative 
example for wind integration, particularly since this produces 
no variability for
the on-the-spot case.  The $f/i$ ratio can be significantly modulated
owing to evolving wind properties.  The extent of the variation depends
on how the time scale for the wind flow compares to the time scale
over which the line emissivities change.
   }
   {
While a variety of factors can illicit variable line ratios,
a time-varying mass-loss rate serves to demonstrate the 
range of amplitude and phased-dependent behavior 
in $f/i$ line ratios.  Importantly,
we evaluate how variable mass loss might bias measures of $f/i$.
For observational exposures that are less than the time scale of
variable mass loss, biased measures (relative to the time-averaged
wind) can result; if exposures are long,
the $f/i$ ratio is reflective of the time-averaged spherical wind.
   }


\maketitle

%
\section{Introduction}
\label{s:introduction}

While our understanding of massive star evolution and the nature
of their stellar winds has advanced tremendously over recent decades,
the advances have themselves generated a swath of new and challenging
questions.  Mass remains the foremost parameter for determining the
destiny of a star \citep{2012ARA&A..50..107L}.  Thus aside from the
many variations that can arise from mass transfer in binary stars
\citep[e.g.,][]{1998A&ARv...9...63V,
2012Sci...337..444S,2014LRR....17....3P}, mass loss can substantially
impact the story line of massive stars
\citep[e.g.,][]{2008A&ARv..16..209P,2014ARA&A..52..487S}.

The most successful theory for wind driving among the early-type
massive stars -- the O stars, early B stars, evolved OB stars, and
even the Wolf-Rayet stars -- is the line-driven wind theory
\citep{1975ApJ...195..157C,1986A&A...164...86P,1986ApJ...311..701F,
1993ApJ...405..738L,1994A&A...289..505S,1995ApJ...454..410G,
1995ApJ...442..296G}.  At the same time, this mechanism also predicts
wind instabilities (i.e., the ``line-driven instability'' mechanism;
hereafter LDI) that lead to the development of shocks
and structured flow \citep{1970ApJ...159..879L,
1980ApJ...241..300L,1988ApJ...335..914O,1997A&A...322..878F}.  While
the LDI is a natural source of structure formation in the wind, it
is also possible that convective processes initiate structure
formation at the wind base \citep{2009A&A...499..279C,
2015ApJ...806L..33A}, without precluding operation of LDI.
Observational support for stochastically structured flow comes in
a variety of forms, including (but not limited to) the black
troughs of ultraviolet (UV) P~Cygni resonance lines
\citep{1983ApJ...274..372L, 1990ApJ...361..607P}, wind clumping
\citep[e.g.,][]{1991A&A...247..455H,1992PhDT........53R,
1998ApJ...494..799E,2003A&A...408..715B,
2006ApJ...637.1025F,2006A&A...454..625P,2008A&ARv..16..209P}, and
of particular interest for this paper the production of X-ray
emissions in the wind
\citep[e.g.,][]{1979ApJ...234L..51H,1981ApJ...250..677C,
1997A&A...322..167B,
2006Ap&SS.304...97S,2006MNRAS.372..313O,2009A&A...506.1055N}.

The ability of {\em Chandra} and {\em XMM-Newton} to provide high
spectral resolution studies of massive star winds has been a major
contributor to further understanding the wind structure
\citep[e.g.,][]{2007A&A...476.1331O,2009A&ARv..17..309G,2013ApJ...770...80L}.
Emission profile shapes of X-ray lines directly probe the kinematics
of the wind flow
\citep{2001ApJ...549L.119I,2001ApJ...559.1108O,2002ApJ...568..954I,
2003A&A...403..217F,2006ApJ...648..565O,2016AdSpR..58..694I} and
can be used to infer mass-loss rates, $\dot{M}$
\cite[e.g.,][]{2014MNRAS.439..908C}.  High resolution spectra have also been
able to resolve, either separately or as partial blends, the triplet
components of He-like species, such as C{\sc v}, N{\sc vi}, O{\sc
vii}, Ne{\sc ix}, and others \citep[e.g.,][]{2007ApJ...668..456W}.  
The three components are referenced
as {\em fir} lines, for forbidden, intercombination, and resonance.
These lines are important because of their diagnostic ability
\cite[e.g.,][]{2001A&A...376.1113P}.  Of
chief interest for this paper is the ratio of line luminosities $R=f/i$.

This line luminosity ratio has a predicted value based on atomic physics, with
different ratio values for different elements.  However, the value can be
modified by pumping effects.  One effect comes from collisional
excitation of the forbidden line that depopulates that level in
favor of the intercombination line \citep{1969MNRAS.145..241G}.
Consequently, for hot plasmas of sufficient density, the line ratio
becomes a diagnostic of the density conditions.  The densities in
massive star winds are generally too low collisional pumping to
be relevant \citep[for an exception, see][]{2017ApJ...845...39O}.
A second process that can change
the line ratio is radiative pumping by UV photons 
\citep{1972ApJ...172..205B}.  Since
massive stars have strong UV stellar radiation fields, and since
the mean intensity of the radiation is a function of distance from
the star (owing to the dilution factor), observed line ratios of
$f/i$ become diagnostics for the vicinity of X-ray producing hot
plasma in relation to the stellar atmosphere 
\citep[early applications
of this diagnostic for massive stars included][]{2001A&A...365L.312K,
2001ApJ...554L..55C}.
The first O~supergiant study with {\it Chandra} HETG data demonstrated the
importance of the 
$f/i$ ratios as a method to establish radial locations in the stellar wind
where He-like 
emission lines form \citep{2001ApJ...548L..45W}.

Several papers have explored the influence of strong stellar radiation
fields on $f/i$ line ratios.
Early applications
made use of an ``on-the-spot'' approximation, whereby the X-rays are
considered to be produced in a shell so that an observed line ratio
could be associated with a single radius in the wind.  Optically thin
X-ray emission from line cooling is a density-squared process, which is
steep in the accelerating portion of the winds, so the assumption that
the X-ray source is dominated by a radially thin shell is a reasonable
zeroth order approximation.  However, the lines are actually formed over
some radial span in the wind, and this will generally be different
in a non-negligible way from the thin shell case.  Integration over
the wind will typically be model-dependent (e.g., the temperature
structure that determines where lines form, the volume filling factor
of the plasma, etc).  \cite{2006ApJ...650.1096L} also found that 
overlapping and wind-broadened lines can influence the strength of
the radiative pumping.

Whether through a shell or wind integration model, work has been
devoted mainly toward understanding line ratios that are not
time-dependent, but there are ways in which the observed ratio can become
time-dependent.  For example, binarity could produce phase-dependent line
ratios by altering photon pumping rates owing to eccentric orbits, or due
to eclipse effects.  Another possibility are co-rotationg interaction
regions (CIRs).  Here the wind of a single star is asymmetric, yet can
be modeled as stationary.  Variations in $f/i$ ratios could arise from
an evolving perspective of a CIR with rotational phase, but will likely
be periodic.  Our focus has been on sources of intrinsic variability
for non-rotating, single stars.  In this case we have split the drivers
for producing time-dependent $f/i$ ratios into two categories:  stellar
variability and wind variability.  Issues of binarity and CIRs are
deserving of separate studies for their impacts on $f/i$ ratios.

Already \cite{2012A&A...542A..71H} explored the first
category in terms of stellar pulsations for modifying the stellar
radiation field to elicit changes in $f/i$ ratios for time-steady winds.
This contribution explores the second category, in which the stellar
radiation is held fixed, but the wind structure is allowed to vary.
Section \ref{s:model} develops our approach based on integrating the line
luminosities throughout the wind.  In particular, we demonstrate that
while the shell approximation is insensitive to changes in the mass-loss
rate, variability in $\dot{M}$ can drive changes in $f/i$ when integration
across the wind is considered.  Section \ref{s:results} provides
illustrative examples.  We explore the extent to which measured line
ratios may be biased in a way that depends on the observational exposure
time.  In section \ref{s:conclusion}, we provide summary remarks and
comments on future work.  An Appendix presents a discussion of effects
from wind attentuation.

\section{Model}
\label{s:model}
%

\subsection{f/i for a Volume Element Source}

Consider an idealized case of a small volume element
in the stellar wind.  This sector of gas has been heated to high
temperature to emit X-rays.  A generic He-like ion is assumed to exist and
to produce a typical {\em fir} triplet emission line.  The line emission
is optically thin; however, the wind may be optically thick to the X-rays.

We introduce the following parameters to
describe the line emission:

\begin{itemize}
\item[--] $L_{\rm f}$ is the total line luminosity in the forbidden component
of the triplet ($^3S_1\rightarrow ^1S_0$ transition).
\item[--] $L_{\rm i}$ is the total line luminosity in the intercombination component
of the triplet ($^3P_{0,1,2}\rightarrow ^1S_0$ transition).
\item[--] $n_{\rm c}$ is the critical number density of electrons for collisional
excitation of electrons into
the intercombination levels out of the forbidden level ($^3S_1\rightarrow ^3P_{0,1,2}$ transition).
\item[--] $\phi_{\rm c}$ is the critical UV photon rate for radiative excitation of
electrons into
the intercombination levels out of the forbidden level ($^3S_1\rightarrow ^3P_{0,1,2}$ transition).
\item[--] $\phi_\ast$ is the stellar UV photon rate.
\item[--] $R_0$ is the ratio $L_{\rm f}/L_{\rm i}$ in the absence of
any pumping processes that depopulate the forbidden level and enhance
the population of the intercombination level.
\item[--] $R$ is the ratio $L_{\rm f}/L_{\rm i}$ that allows for 
pumping effects.
\end{itemize}

Assuming electron densities are relatively small such that $n_{\rm
e} \ll n_{\rm c}$, a well-known result for the ratio of line emission
is \citep[e.g.,][]{2001A&A...365L.312K}:

\begin{equation}
R = \frac{dL_{\rm f}/dV}{dL_{\rm i}/dV} = \frac{R_0}{1+2k_\ast\,W(r)},
	\label{eq:Rvol}
\end{equation}

\noindent where $dL_{\rm f}/dV$ and $dL_{\rm i}/dV$ represent the
luminosity contributions from the volume element,
$k_\ast = \phi_\ast/\phi_{\rm c}$, and $W$ is the dilution
factor given by

\begin{equation}
W(r) = \frac{1}{2}\,\left[ 1- \sqrt{1-\frac{{\cal R}_\ast^2}{r^2}}\right],
\end{equation}

\noindent for $r$ the radius in the wind and ${\cal R}_\ast$ the stellar radius.
Equation~(\ref{eq:Rvol}) for $R$ assumes the volume element has a constant 
temperature.  However, the ratio $R_0$ is weakly dependent on
temperature, and so the same formula may be used in an approximate way
even for a multi-temperature
plasma.  

\subsection{f/i Ratio for a Shell Source}

Instead of a volume element, now consider 
a thin spherical shell, of width $dr$.  Imagine the shell is traveling through
the wind following a velocity profile, $v(r)$.  To determine the line ratio,
contributions to $L_{\rm f}$ and $L_{\rm i}$ must be accumulated for
the unresolved shell, with

\[ \frac{dL_{\rm f}}{dr} = 2\pi\,\int^{+1}_{\mu_\ast(r)}\,r^2\,\frac{dL_{\rm f}}
	{dV}\,d\mu, \]

\noindent and 

\[ \frac{dL_{\rm i}}{dr} = 2\pi\,\int^{+1}_{\mu_\ast(r)}\,r^2\,\frac{dL_{\rm i}}
	{dV}\,d\mu, \]

\noindent where $\mu=\cos \theta$ for $\theta$ the polar angle from
the observer's line-of-sight, $\mu_\ast(r) = \sqrt{1-{\cal R}_\ast^2/r^2}$ accounts for
stellar occultation of part of the shell when at radius $r$ in the wind,
and due to symmetry
integration in the azimuth $\phi$ about the observer's axis
has already been carried out.

For a thin spherical shell, the two integrals yield a result that is
the same as for a small volume element, with

\begin{equation}
R = \frac{dL_{\rm f}/dr}{dL_{\rm i}/dr} = \frac{R_0}{1+2k_\ast\,W(r)}.
	\label{eq:Rshell}
\end{equation}

\noindent Given that $R_0$ is a constant, the only way to produce
variability in the ratio $R$ is either for $k_\ast$ to change, or
for the location of the shell, $r$, to change.  \cite{2012A&A...542A..71H}
explored the possibility of
a time-dependent $R$ being driven by variability in the stellar radiation
field.
Here, the focus is on factors that
alter $R$ owing to wind structure.

For a simple spherical shell, ignoring the temperature influence
(i.e., assuming $R_0$ is fixed and that the hot plasma has a 
temperature adequate to produce the line emission under
consideration), variations in $R$ naturally arise as the shell
evolves through the wind.  For illustrative purposes, consider
a shell that is coasting at constant speed with $v(r) = v_0$.
After a time-of-flight $t$, with the shell originating at the stellar
surface, the
radial location of the shell becomes

\begin{equation}
r(t) = {\cal R}_\ast + v_0\,t.
\end{equation}

\noindent Time-dependence in the line ratio $R$ enters through the
dilution factor.  The dilution factor ranges from 0.5 (at $r={\cal
R}_\ast$) to 0.0 (as $r\rightarrow \infty$), hence $2W$ ranges between
0 and 1.  At large distance, $W \propto r^{-2}$.  If $k_\ast \gg 1$,
the shell may have to travel great distance before $R$ changes.

As a more realistic case, the velocity profile of a stellar
wind is frequently approximated as a beta-law, with

\begin{equation}
v(r) = v_\infty\,(1-bu)^\beta,
\end{equation}

\noindent where $u={\cal R}_\ast/r$ is a normalized inverse radius, and $b$ is
a constant that serves to set the initial wind speed at the wind base,
with $v_0 = v_\infty\,(1-b)$.
For use as an example, we introduce a normalized velocity with $\beta=1$:

\begin{equation}
w(u) = v/v_\infty = 1-bu.
\end{equation}

\begin{figure}
\includegraphics*[width=\columnwidth]{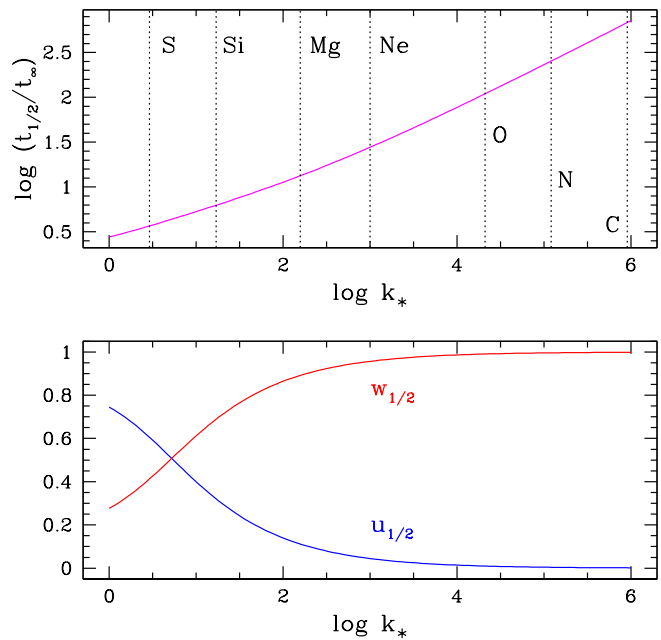}
\caption{For an X-ray emitting shell moving through a wind
with a $\beta=1$ velocity law, the two panels show the characteristic
time, distance, and velocity for which the $R=f/i$ line ratio will change.
Upper panel is for the characteristic time $t_{1/2}$, relative to the wind
flow time, $t_\infty$.  Lower panel is for the distance and velocity
in the wind corresponding to $t_{1/2}$.  See text for explanation
of the vertical lines in the upper panel.
}
\label{fig1}
\end{figure}

\noindent Figure~\ref{fig1} illustrates the characteristic
time over which $R$ will vary as a geometrically thin shell moves through
the wind following the velocity law, $w(u)$.  Note that at large
distance, $R \rightarrow R_0$, and the line ratio has a minimum of
$R_{\rm min} = 1/(1+k_\ast)$ when the shell is at $r={\cal R}_\ast$.
We define $t_{1/2}$ as the time-of-flight for the shell until $R
= 0.5(R_{\rm min}+R_0)$.

The upper panel of Figure~\ref{fig1} gives $t_{1/2}/t_\infty$, where
$t_\infty={\cal R}_\ast/v_\infty$ is the characteristic flow time for
the wind, against $k_\ast$ in a log-log plot.  Vertical lines are for
different He-like ion species assuming a stellar radiation field for
$\zeta$~Pup, using a Kurucz model with $T_{\rm eff} = 40,100$~K
and $\log g = 3.65$ \citep{2001ApJ...554L..55C}.  For a different star,
the vertical lines would shift laterally for the appropriate radiation
field at the stellar surface.  Note that for massive star winds,
$t_\infty$ is of order hours or a day.  What the upper panel shows is
that different lines will tend to have different response times for how
$R$ varies.  The lower panel shows where in inverse radius, $u_{1/2}$,
or in normalized velocity, $w_{1/2}$, the ratio $t_{1/2}/t_\infty$
is achieved as a function of $k_\ast$.

Before exploring the line ratio based on integration throughout the
wind, it it is worth noting here that for a shell at a fixed location,
the line ratio is insensitive to a time variable wind density.
While the emission in all of the triplets will change with density,
they all rise or fall by the same factor for X-ray emission produced
at a fixed distance from the star.

\subsection{f/i Ratios from Distributed Sources}

\cite{2006ApJ...650.1096L}
presented an approach for evaluating the emission line ratio when
the hot plasma is distributed throughout the wind.  Assuming spherical
symmetry, and optically thin line emission, the emissivities for
the forbidden and intercombination components are given by

\begin{equation}
j_{\rm f} \propto \frac{\tilde{R}}{1+\tilde{R}}\,\rho^2\,
\end{equation}

\noindent and

\begin{equation}
j_{\rm i} \propto \frac{1}{1+\tilde{R}}\,\rho^2\,
\end{equation}

\noindent where $\rho(r)$ is the mass density of the hot plasma in
the wind, and 

\begin{equation}
\tilde{R}(r) = \frac{dL_{\rm f}/dr}{dL_{\rm i}/dr} 
	= \frac{R_0}{1+2k_\ast\,W(r)}.
\end{equation}

\noindent Since $R$ is the notation for the observed line ratio,
the addition of a tilde in the above merely signifies the
ratio for just one shell in the wind in which hot plasma is distributed
over a range of radii.
The luminosities in the respective lines are given by

\begin{equation}
L_{\rm f,i} \propto \int\,j_{\rm f,i}\,(1-W)\,r^2\,dr,
\end{equation}

\noindent where the parenthetical involving the dilution factor 
accounts for the effect of stellar occultation.

It is possible to recast the line ratio to mimic somewhat the classic
result for a shell in equation~(\ref{eq:Rshell}).  Using inverse radius
$u={\cal R}_\ast/r$, we begin as follows:

\begin{equation}
\frac{1}{1+\tilde{R}} = \frac{1+2k_\ast\,W}{A_0-k_\ast\,\sqrt{1-u^2}},
\end{equation}

\noindent where 

\begin{equation}
A_0 = 1 + R_0 + k_\ast.
\end{equation}

\noindent Then

\begin{equation}
\frac{\tilde{R}}{1+\tilde{R}} = \frac{R_0}{A_0-k_\ast\,\sqrt{1-u^2}}.
\end{equation}

\noindent With these conversions, using the normalized wind
velocity $w=v/v_\infty$, and inserting $\rho \propto u^2/w$, the luminosity
for forbidden line emission is

\begin{equation}
L_{\rm f} = L_0\,R_0\,\int\left[ \frac{1}{A_0-k_\ast\,\sqrt{1-u^2}}\right]\,
	(1-W)\,\frac{du}{w^2} \equiv L_0\,R_0\,\Lambda_0,
\end{equation}

\noindent where $L_0$ is a constant that will cancel when taking
the line ratio.  Next,

\begin{eqnarray}
L_{\rm i} & = & L_0\,R_0\,\int\left[\frac{1+2k_\ast\,W}{A_0-k_\ast\,
	\sqrt{1-u^2}}\right]\,(1-W)\,\frac{du}{w^2} \nonumber \\
 & = & L_0\left\{ \frac{L_{\rm f}}{R_0}(1+k_\ast) - k_\ast\int \left[
	\frac{\sqrt{1-u^2}}{A_0-k_\ast\sqrt{1-u^2}}\right](1-W)\,
	\frac{du}{w^2} \right\}\nonumber \\
 & \equiv &  L_0\,\Lambda_0\,(1+k_\ast) - L_0\,k_\ast\,\Lambda_1.
\end{eqnarray}

\noindent Note that in all of the preceding integrals, the
upper and lower limits are formally for the radial intervals over
which the line in question forms.  In principle, there could be
multiple such radial zones, and their locations and spatial extents
could be functions of time.  

The line ratio, now involving all of the forbidden and intercombination
line emission separately evaluated throughout the wind, becomes

\begin{equation}
\frac{L_{\rm f}}{L_{\rm i}} = R =\frac{R_0\Lambda_0}{(1+k_\ast) 
	\,\Lambda_0-k_\ast\,\Lambda_1}.
\end{equation}

\noindent Finally, one can recast this relation as

\begin{equation}
R \equiv \frac{R_0}{1+k_\ast\,(1-\xi)},
	\label{eq:Rwind}
\end{equation}

\noindent where $\xi = \Lambda_1/\Lambda_0$, and the overall expression
bears strong similarity to equation~(\ref{eq:Rvol}) with $1-\xi$
acting in the place of $2W(r)$.  For $r\rightarrow \infty$, $\xi
\rightarrow 1$.  The minimum value of $\xi$ for $r={\cal R}_\ast$
will depend on line-specific parameters, but can be as low as zero.
All of the effects of wind integration are collected in the parameter
$\xi$.  This parameter also depends on factors that are specific to the
line under consideration, with $\xi = \xi(R_0, k_\ast, u_{\rm min},
u_{\rm max})$, 
and $u_{\rm min}$ and $u_{\rm max}$ are limits for the wind integration
that are set by where X-ray production occurs, or by specifics
of the temperature distribution relevant to the line in question.
As a result, $\xi$ differs from one triplet to the next.

Note that the expressions above for wind integration can also take
into account photoabsorption of X-rays by the wind. The inclusion of
this effect does not alter the key result of equation~(\ref{eq:Rwind}).
Photoabsorption introduces an exponential factor of wind optical
depth in the respective integrands for the forbidden and intercombination
line emissions.  With wind absorption 
the two line luminosity expressions become

\begin{equation}
L_{\rm f} = L_0'\,R_0\,\int\int\left[\frac{1}{A_0-k_\ast\,\sqrt{1-u^2}}\right]\,
	(1-W)\,e^{-\tau(u,\mu)}\,\frac{du}{w^2}\,d\mu ,
\end{equation}

\noindent and

\begin{equation}
L_{\rm i}  =  L_0'\,R_0\,\int\int\left[\frac{1+2k_\ast\,W}{A_0-k_\ast\,
	\sqrt{1-u^2}}\right]\,(1-W)\,e^{-\tau(u,\mu)}\,\frac{du}{w^2} \,d\mu,
\end{equation}

\noindent where the wind optical depth is

\begin{equation}
\tau(u,\mu) = \int\,\kappa\,\rho\,dz.
\end{equation}

\noindent The coefficient changes with $L_0 \rightarrow L_0'$ because
the integration is now a double integral in both inverse radius $u$
as well as polar angle in the form of $\mu$, since the optical depth
is evaluated along a ray of fixed impact parameter.  Many authors
have presented ways of handling wind absorption \citep[e.g.,][]
{2010ApJ...719.1767L}.  The Appendix provides further discussion on
the topic.  In what follows the wind absorption will be ignored for the
sake of example cases.

\section{Model Results}
\label{s:results}

Wind integration enlarges the possibilities for variability not just
in the separate emission lines of the triplet, but in the ratio $R$
as well.  Focusing strictly on drivers of variability from changes in
the wind (i.e., ignoring changes in $k_\ast$), factors that could induce
variability in $R$ include changes in the wind density and changes in the
temperature distribution.  For the wind density, global changes to the
wind might include the mass-loss rate $\dot{M}$, or the wind velocity law.
Time dependence in any of $\dot{M}$, $v_\infty$, $b$, or $\beta$ would
lead to time dependence in $\rho$.  This is a particularly interesting
result, since a shell model has no sensitivity to density variations.
Time dependence of the temperature distribution for the hot plasma will
influence $R$ as well.  This can arise from changes in the range of
temperatures achieved in the wind, or the radial profile of the distribution.

However, whether in the density or in the temperature distribution, 
creating an observable $R(t)$ will mainly result if there
is a global change in the wind, as opposed to distributed and stochastic
changes that produce effectively a steady-state wind.

As an illustrative
case, we consider a sinusoidal variation in the mass-loss rate.
This variation is modeled with

\begin{equation}
\dot{M}(t) = \dot{M}_0 + \delta\dot{M}\,\sin\left(\omega t-\Phi_0\right),
	\label{eq:mdot}
\end{equation}

\noindent where $\dot{M}_0$ is the average value of the mass-loss rate,
$\delta\dot{M}$ is an amplitude for the variation, $\omega =
2\pi/P$ is the angular frequency for period $P$ associated
with the cyclical variation, and $\Phi_0$ is an arbitrary phase.
Note that $\delta\dot{M} \le \dot{M}_0$, otherwise density would
become negative.  

However, equation~(\ref{eq:mdot}) is how the mass loss varies
at the base of the wind.  The density perturbation elsewhere
depends on the flow time between the base and the radius of interest.
To determine the density at all radii in the wind, one must determine
the time lag between the radius $r$ and the surface ${\cal R}_\ast$.
This time lag, $t_{\rm lag}$, is the time of travel through
the wind.  For the sake of illustration, we assume the velocity is
a $\beta=1$ velocity law.  
The time lag then becomes

\begin{equation}
t_{\rm lag} = t_\infty\,\left\{\frac{1-u}{u}+b\,\ln\left[\frac{1-bu}
	{(1-b)\,u}\right]\right\} \equiv t_\infty\,\gamma(u),
\end{equation}

\noindent where again $t_\infty = {\cal R}_\ast/v_\infty$ is the
characteristic flow time in the wind.  
Now the mass-loss rate at
any location in the wind at any time is

\begin{equation}
\dot{M}(u,t) = \dot{M}_0 + \delta\dot{M}\,\sin\left\{2\pi\,\left[
	\frac{t}{P}-\frac{t_\infty}{P}\,\gamma(u)\right]-\Phi_0\right\}. 
\end{equation}

\noindent The density is given by

\begin{equation}
\rho(t,u) \propto \frac{\dot{M}(t,u)}{w(u)}\,u^2.
\end{equation}

\noindent The formulations for $\Lambda_0$ and $\Lambda_1$ are 
unchanged, except they now become functions of time following
the integration over volume, because the density undulates as
a propagating wave.  

The result for the line ratio is

\begin{equation}
R(t) = \frac{R_0}{1+k_\ast\,[1-\xi(t)]}.
\end{equation}

\begin{figure}
\includegraphics*[width=\columnwidth]{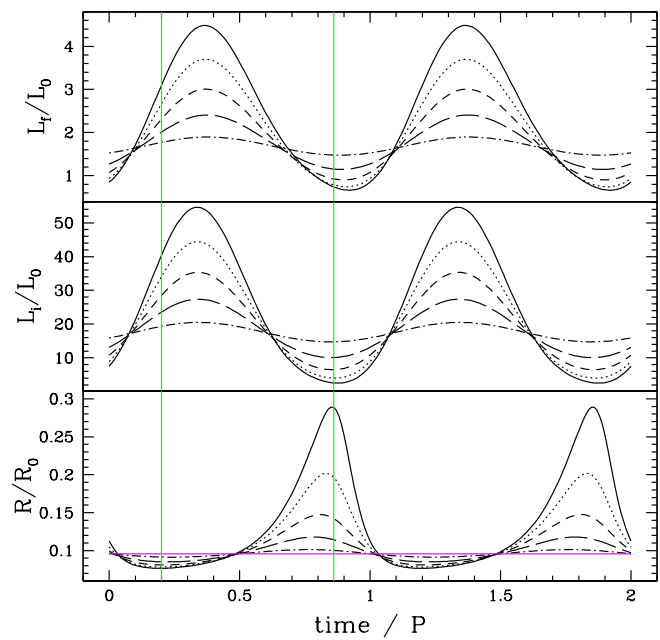}
\caption{
A plot of variability in the forbidden line luminosity (upper),
intercombination line luminosity (middle), and the line ratio
(lower).  Here, $L_0$ is a constant, and $R_0$ is the line ratio
in the absence of UV pumping.  These are plotted against time
relative to the period $P$ for variability in the mass-loss rate.
The different curves are for $\delta\dot{M}/\dot{M}_0 =$ 0.1
(dash-dot), 0.3 (long dash), 0.5 (short dash), 0.7 (dotted),
and 0.9 (solid).  In the lower panel, the horizontal line in
magenta is for $\delta\dot{M}=0.0$.  The two vertical green 
lines are for the minimum and maximum values of $R/R_0$ when
$\delta\dot{M}/\dot{M}_0 =0.9$.  Two cycles of the periodic
variability are shown for clarity of viewing.
}
\label{fig2}
\end{figure}

\noindent Examples of $R(t)$ plotted with phase for cyclic
variability in the mass loss are shown in Figure~\ref{fig2}, with
two cycles shown for better display of the variation.  The different
curves are for different values of $\delta \dot{M}/\dot{M}_0$.  
At top is the relative luminosity in the forbidden line.  Middle
is for the intercombination line.  Bottom is the line ratio
relative to $R_0$.
In the lowest panel, the
horizontal line in magenta is the result when mass loss is constant
(i.e., $\delta \dot{M} = 0$).   The vertical green lines are for
the minimum and maximum in $R$ for $\delta \dot{M}/\dot{M}_0=0.9$,
as a specific example allowing comparison between the extrema
in the line ratio with the individual line luminosities.
All of the curves are for $\Phi_0=0$.

For this Figure we adopt parameters similar to $\zeta$~Pup as
a general guideline.  We also used a ratio of $t_\infty/P = 0.1$.
The value of $\xi$ depends on the choice of
$R_0$; a value of $R_0=2.5$ for Si{\sc xiii} was used for this example
\citep{1972ApJ...172..205B}.  We again adopt the stellar radiation
field used by \cite{2001ApJ...554L..55C} for $\zeta$~Pup, which gives
$k_\ast = 16.9$ for the rate of UV pumping of the $f$-line component at
the stellar surface.  For the volume integration of line luminosities,
the X-rays are considered to exist from the stellar surface ($u_{\rm
max}=1$) to infinity ($u_{\rm min}=0$).

There are several features worth noting about Figure~\ref{fig2}.
To begin, the peaks are much taller than the troughs are deep.
The emissivity scales with the square of density.  As a result,
a snapshot of the wind reveals alternating over- and under-density
zones relative to the mean.  With a $\rho^2$ emissivity, the series of
shells act much like a clumped wind.  But since the UV pumping is
strongest at the inner wind, the presence of an increased density at
locations where the UV pumping is diminished can considerably enhance
the luminosity in the forbidden line, relative to a time-steady flow.
The intercombination line is enhanced where pumping is strong, so the
demoninator for the line ratio is also changing.  The vertical green
lines are guides to aid in comparing the state of the respective line
luminosities to the varying line ratio.

\begin{figure}
\includegraphics*[width=\columnwidth]{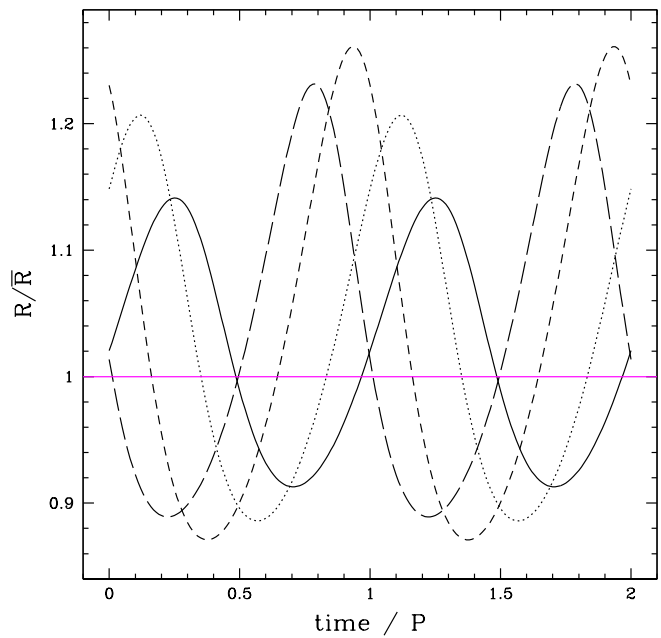}
\caption{Shown is a comparison of $R=f/i$ line ratios when the
line forms beyond radius $r_0$, with $r_0/{\cal R}_\ast=
1.0$ (long dash), 1.1 (short dash), 1.5 (dotted), and 2.0 (solid),
with $\delta\dot{M}/\dot{M}_0=0.3$.  $R$ is normalized to
$\bar{R}$, which is the value for the line ratio when $\delta\dot{M}
=0$ for the respective cases.  Thus the curves are relative changes
to a non-varying wind, indicated as the horizontal line in
magenta.
}
\label{fig3}
\end{figure}

In Figure~\ref{fig2} the emission is assumed to form from the wind
base at ${\cal R}_\ast$ to infinite distance (although emission at
quite large radius will have minimal contribution for the line flux).
However, many studies treat the inner radius for the production of X-rays
as a free parameter for model fits.  For example, in a line profile
analysis of several lines measured by {\em Chandra} for $\zeta$~Pup,
\cite{2010MNRAS.405.2391C} found that X-rays were produced from
$r_0=1.5{\cal R}_\ast$ and beyond.  Figure~\ref{fig3} compares examples
with values of $r_0/{\cal R}_\ast=1.0$ (long dash), 1.1 (short dash),
1.5 (dotted), and 2.0 (solid), all with $\delta\dot{M}/\dot{M}_0 = 0.3$.
The cases have quite different values for $R$ in the absence of variable
mass-loss; consequently, Figure~\ref{fig3} displays a relative variation
for ease of comparison, where each case is normalized to $\bar{R}(r_0)$
as the value for its non-varying wind.  The shapes are generally similar,
although phase shifted owing to time lags for the flow traversing the gap
$r_0-{\cal R}_\ast$.  The relative peak-to-trough amplitudes are actually
non-monotonic with $r_0$, but ultimately drops as $r_0$ increases to
larger values, as the radiative pumping becomes weaker with distance.

\begin{figure}
\includegraphics*[width=\columnwidth]{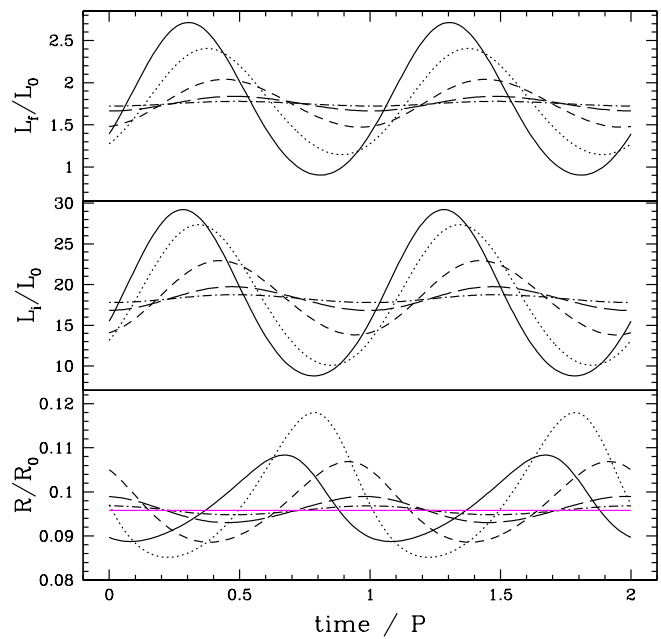}
\caption{Similar to Fig.~\ref{fig2} but with 
$\delta\dot{M}/\dot{M}_0=0.3$ and
different ratios of $t_\infty/P=0.03$ (solid), 0.1 (dotted), 0.3
(short dash), 1.0 (long dash), and 3.0 (dash-dot).
}
\label{fig4}
\end{figure}

Another consideration is to vary the ratio $t_\infty/P$, with a selection
of examples displayed in Figure~\ref{fig4}.  As in Figure~\ref{fig3},
$\delta\dot{M}/\dot{M}_0 = 0.3$ is held fixed.  The three panels follow
those of Figure~\ref{fig2}, now with $t_\infty/P=0.03$ (solid),
0.1 (dotted), 0.3 (long dash), 1.0 (short dash), and 3.0 (dash-dot).
Altering the period at which $\dot{M}$ is varied relative to the wind flow time
leads to both phase shifting in the pattern and amplitude changes.
The amplitude of variation in $R$ drops as $t_\infty/P$ increases.
For a wind with relatively high frequency oscillations
in $\dot{M}$, the wind density varies over short length scales, and
the wind integrations for $L_{\rm f}$ and $L_{\rm i}$ obtain values for the
time-averaged stationary wind.

Returning to Figure~\ref{fig2}, perhaps the most important point is
that these curves are for a {\em snapshot} of the wind, as if
measures for $L_{\rm f}$ and $L_{\rm i}$ were instantaneous.  
However, single massive
stars are relatively faint X-ray sources.  The exposures
required to obtain sufficient counts for high signal-to-noise line
fluxes with current facilities is measured in many kiloseconds
of data collection.

\begin{figure*}[t]
\begin{centering}
\includegraphics[width=1.7\columnwidth]{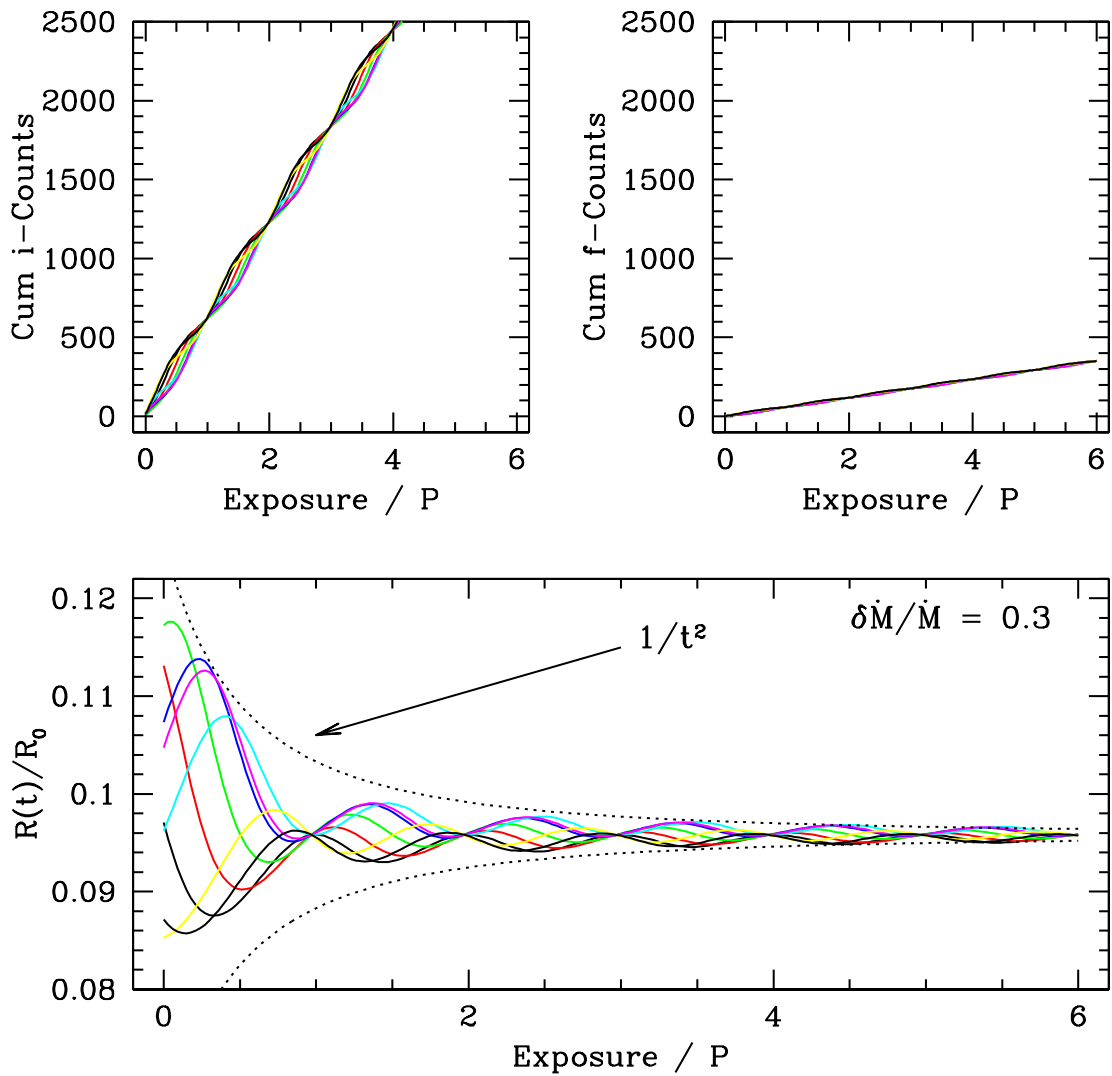}
\caption{The figure shows the effect of exposure time for
a hypotehtical observation of a He-like triplet.  Upper left
shows the cumulative counts in the i-component of
the triplet with exposure time, as normalized to the period
for variability in the mass-loss rate.  Upper right is
for the cumulative counts in the f-component.  Bottom
is for the line ratio.  The different curves are for
different phase values, $\Phi_0$ (see text).  The
example is for $\delta\dot{M}/\dot{M}_0 = 0.3$ for
Si{\sc xiii} with stellar parameters for $\zeta$~Pup.
Also shown in the lower panel in dotted line type is
a $t^{-2}$ decay envelope.
}
\label{fig5}
\end{centering}
\end{figure*}

In practice one is not concerned so much about the variable luminosity
of the forbidden and intercombination lines so much as the accumulated
counts (or energy) over the course of an exposure.  Figure~\ref{fig5}
shows how exposure time affects the measured value of the line ratio
that includes the time-varying wind density.
The top two panels plot the cumulative counts in the i and f lines,
respectively, against the exposure time as normalized to the period for
the variation in mass loss.  The relative scale is arbitrary, since what
matters for $R$ is the ratio of the two line luminosities.

If the X-ray luminosities in the lines were constant, the counts would
grow linearly in time.  However, the variability in the mass-loss rate
imposes the wavy structure seen on the otherwise linear growth in both
the i and f line counts.  The different colors are for eight different
phases, with $\Phi_0 = 0^\circ, 45^\circ, 90^\circ, 135^\circ, 180^\circ,
225^\circ, 270^\circ,$ and $315^\circ$.

The lower panel shows the line ratio as a function of the exposure
time for the observation, again in terms of the period of variability
for $\dot{M}$.  The curves are for $\delta{M}/\dot{M}_0 = 0.3$.
Figure~\ref{fig5} shows that if the period is long compared to
the exposure, then the line counts and the resulting line ratio $R$
basically reflect a snapshot for the current state of the wind's inner
density.  However, if the exposure time is relatively long, then the
oscillatory perturbation in the wind becomes diminishingly relevant in
terms of the accumulated counts.  As a result, the line ratio $R$ 
achieves a value for the wind as if it were non-varying,
at the time-averaged mass-loss rate $\dot{M}_0$.  Since the emissivity
scales as $\rho^2$, a roughly $t^{-2}$ decline in the variation of $R$
relative to a non-varying wind results; overplotted is an envelope for
a decline in amplitude with $t^{-2}$ shown a dotted lines.

It is useful to explore observational prospects for detecting effects
displayed in Figure~\ref{fig5}.  Since parameters for $\zeta$~Pup were
adopted in the examples, existing data for this star are considered.
While considerable data have been obtained with the {\it XMM-Newton}
\citep[e.g.,][]{2018A&A...609A..81N}, the various datasets are
spread over many years and not suitable for this type of study.
\cite{2001ApJ...554L..55C} obtained 67~ks of continuous high-resolution
{\em Chandra} data for $\zeta$~Pup.  This corresponds to an exposure of
about 19~hours.  Using radius and terminal speed values from that paper,
the flow time is $t_\infty =1.6$~hours, and the {\em Chandra} exposure
is nearly 12 flow times, not far from the value of 10 flow times used in
our examples.  \cite{2001ApJ...554L..55C} measured $R=1.04\pm 0.14$ for
Si{\sc xiii}, which is a 13\% uncertainty in $R/R_0$.  Figure~\ref{fig5}
suggests that $R/R_0$ might be 5\% higher or lower than the long-term
value.  One would need multiple pointings of a similar duration at an
accuracy of around 1\% to detect the envelope of variability, assuming
$\delta \dot{M}/\dot{M} = 0.3$.   This would allow the creation of a
scatter plot of values from the mulitple pointings, that could be compared
with a diagnostic plot like Figure~\ref{fig5}.  Assuming variability is
not regular in the long-term (e.g., repeating with phase), the ensemble of
measures should fill in the envelope of possibilities, namely 
between the $t^{-2}$
pair of curves.  Existing data do not appear adequate to the purpose.
Of course, larger values of $\delta \dot{M}/\dot{M}$ would be easier
to detect, while smaller variations would be harder.

\section{Conclusion}
\label{s:conclusion}
%

In this paper we have explored drivers of variable $R=f/i$ line luminosity
ratios for the wind from UV-bright massive stars, with application
to variable mass-loss as a quantitative example.  Whereas a previous
paper described the influence of a variable stellar radiation field
\citep{2012A&A...542A..71H}, here the focus has been on variability within
the wind itself.  This variability can arise from altering anything that
can change the emissivity of line production (e.g., temperature
structure or density structure).  
However, in order to achieve significant variations in $R$, the
variation in the wind must be global in nature.  Small-scale stochastic
variations will not much engender time-dependence in the line ratio $R$.

As proof-of-concept, we considered a variable mass-loss rate, taken
simply as a sinusoid in time with period $P$.  The X-ray emissivity for
line cooling is density squared, and so the smoothly undulating density
with radius affects the line emission much as clumping in the form of
spherical shells.  This affects both the forbidden and intercombination
lines equally in terms of density; however, the UV pumping of the
forbidden line population serves as an additional radius-dependent
weighting factor for determination of the respective line luminosities.
As a result, a periodic, but non-sinusoidal, variation in $R$ persists.

Observations are obtained over relatively long exposure times.
Allowing for the accumulation of line counts over time shows that the
relevance of variable $f/i$ ratios will depend on how the exposure time
for the observations compare to the period of the variable mass loss.
If the exposure time is short compared to $P$, then the $f/i$ ratio may be
biased in terms of the phase of $\dot{M}(t)$ at which data were obtained.
An analysis based on a steady-wind model would thus lead to errors in the
distribution of the hot plasma, in relation to the stellar atmosphere.
If the exposure is long, the effects of time-varying wind density averages
out in the accumulation of line counts, and the measured value of $R$
will obtain a value representing the time-averaged spherical wind.

In practice, long observation times are not generally achieved in a single
continuous exposure.  Instead, a source may be visited multiple times,
each one a subexposure for the program.  These subexposures may
not be of equal duration nor equally spaced in time.  They will likely
be obtained over the course of a year's span, since most programs run
on an annual cycle.  While we did not conduct simulations for ensembles
of subexposures, the logic above still applies.  If the duty cycle of
the subexposures is long compared to $P$, then the wind is effectively
randomly sampled.  But if $P$ is long, then the various subexposures
essentially sample a relatively fixed phase for the wind.  Understanding
how observed $f/i$ ratios may be biased for the inbetween cases would
require further model simulations.

We note that \cite{2018MNRAS.481.5263D} have presented
results for 1D time-dependent hydrodynamical simulations for line-driven
winds with a sinusoidally varying stellar radiation field on a period
$T_S$ (``S'' for source).  We can associate our period, $P$, for the
varying $\dot{M}$ with their period notation.  They then introduce a
dynamical time as a ratio of the radius for the wind critical point,
$r_{\rm c}$, to the flow speed at the critical point, $v_{\rm c}$.
Supposing $r_{\rm c}= {\cal R}_\ast$, and $v_{\rm c}=v_0$, the latter
being the wind speed at the base of the wind, then in terms of our flow
time, their dynamical time becomes $t_{\rm c} = (v_\infty/v_0)\,t_\infty
\sim 10^2\,t_\infty$.  \cite{2018MNRAS.481.5263D} find that for $P \gg t_{\rm
c}$, the wind oscillates between so-called high and low states, meaning
the wind mass loss reflects the state of the stellar radiation field.
When $P \ll t_{\rm c}$, the wind is largely stationary as if driven
by a constant radiation flux (i.e., the average radiation field of
the star).  All of our examples in this paper are in the long-period
regime of \cite{2018MNRAS.481.5263D}, since even $t_\infty/P = 3$
(see Fig.~\ref{fig4}) corresponds only to $t_{\rm c}/P \sim 0.03$.

While \cite{2012A&A...542A..71H} considered the effects of a variable
radiation field for producing variability in $f/i$ line ratios, and this
paper has emphasized the effects of variable wind structure, the two
may well be linked.  The examples here were limited to
fluctuating mass loss.  To explore how $f/i$ ratios could be
impacted when both $\dot{M}$ and the stellar luminosity $L_\ast$
are time-dependent, we consider a
line-driven wind: \cite{1999isw..book.....L} state that $\dot{M}
\sim L_\ast^{1.5}$, which implies $\delta \dot{M}/ \dot{M} \sim 1.5\,
\delta L_\ast/L_\ast$.  Assume that the stellar
luminosity variations occur for a star of fixed radius, then $\delta
L_\ast/L_\ast = 4\delta T_\ast/T_\ast$.  In the hottest star considered
by \cite{2012A&A...542A..71H} ($T_\ast = 40,000$~K, similar to that of
$\zeta$~Pup and the examples of this paper), the wavelength for
pumping associated with Si{\sc xiii} is approximately
in the Rayleigh-Jeans tail of
the blackbody, and thus linear in $T_\ast$, implying that $k_\ast \propto
T_\ast$ for this scenario.  It also implies that the variable mass loss
and luminosity are in phase.  With $\delta T_\ast/T_\ast = (1/6)\delta
\dot{M}/ \dot{M}$, and $\delta \dot{M}/ \dot{M}=0.3$, we found that the
range of variation of $R/R_0$ in Figure~\ref{fig2} increased by less than
5\%, thus the response to phased luminosity changes is less than linear.
However, this example artificially fixes the stellar radius, does not take
account of non-radial pulsations, and considers only the Rayleigh-Jeans
limit.  The extent to which different lines respond to both wind structure
and variable luminosity is an interesting study for a future paper.

\begin{acknowledgements} 
We are grateful to an anonymous referee for comments that have improved
this manuscript.  We thank Wayne Waldron for his support and insights
concerning the topic of this study.  This research was supported by NASA
grant G08-19011F for the Chandra General Observer Program, Cycle 19.
\end{acknowledgements}

\bibliographystyle{aa}     
\bibliography{ignace}

\begin{appendix}

\section{Effects of Wind Attenuation on $f/i$ Line Ratios}

Some winds are sufficiently dense that photoabsorptive opacity
suppresses the escape of X-rays from the wind and influences the
ionization balance in the wind \citep[e.g.,][]{1992A&A...266..402B,
2010ApJ...711L..30W, 2016AdSpR..58..710K}.  The effect can depend
on abundances, owing to the large cross-sections of metal ions
\citep[e.g., solar versus metal-rich such as Wolf-Rayet stars,
see][] {1999A&A...348L..45I}.  The cross-section scales roughly as
cube of the wavelength,
$\lambda^3$, so the strength of photoabsorption ranges substantially
across an X-ray spectrum, from being significant at soft energies
to potentially irrelevant at high ones.  Since radiative pumping is strongest
for X-rays formed closest to the star, and since photoabsorption
(when relevant) will absorb the innermost X-rays of the
wind, attenuation effects can impact the observed $f/i$ line ratios
\citep[e.g., the observed $f/i$ line ratios for WR~6, with a quite dense
wind, are consistent with no UV pumping;][]{2015ApJ...815...29H}.
In terms of variability, attenuation modifies where in the
wind X-rays can escape to the observer, and as indicated in the
discussion for Figures~\ref{fig1} and \ref{fig3}, location determines
the timescale and amplitude of variability in $f/i$ ratios.

The inclusion of photoabsorption to determine emergent line luminosities
throughout the wind is straightforward, with

\begin{equation}
L_{\rm f,i} = \int_{{\cal R}_\ast}^\infty\,\int_{\mu_0(r)}^{+1}\,
	2\pi\,j_{\rm f,i}\,e^{-\tau(r,\mu)}\,r^2\,dr\,d\mu,
\end{equation}

\noindent where $\mu_0(r)$ accounts for the effect of stellar
occultation, and the optical depth to photoabsorption is

\begin{equation}
\tau(r,\mu,\lambda) = \int \kappa(\lambda)\,\rho(r)\,dz,
\end{equation}

\noindent The opacity $\kappa(\lambda)$ does not vary much between triplet
components of a given He-like species, but can change significantly
from the triplets of one species
to the next.  The density $\rho$ refers to the ``cool'' component
(not X-ray producing), which produces features such as
UV P~Cygni lines.  The optical depth is for a ray of fixed impact parameter,
hence the integration in $z$.  

Without photoabsorption, the line luminosity calculation is a 1D integral
in radius; with it, the evaluation is a 2D integral.  However,
a 1D integral can be recovered using the exospheric approximation
\citep[example applications
of the exospheric approximation for stellar wind X-rays include][]
{1999ApJ...520..833O,2000MNRAS.318..214I}.
The exospheric
approximation does not produce quantitatively accurate results;
however, it can produce qualitatively accurate trends, and so here
we employ it for heuristic purposes.

The approximation is to determine the radius along the line-of-sight
to the star where $\tau(E)=1$, denoted as $r_1(E)$.  This radius is treated
as a hard spherical boundary for which no X-rays escape when $r<r_1$,
and X-rays formed at $r>r_1$ escape without attenuation.  The
occultation factor $\mu_0$ is modified for what is effectively a
wavelength-dependent stellar size.

The upshot for calculation of line luminosities is that the lower limit
for the integration in radius (or the upper limit in terms of inverse
radius) is the greater of $r_1$ and $r_0$.  In the illustrative
case of variable mass-loss explored in this paper, the emissivity
scales as density squared, whereas photoabsorption optical depth
is only linear in density, yet the attenuation is exponential in
the optical depth.  Ultimately, larger values of $r_1$ will
tend to drive the line ratio to $R\rightarrow R_0$, and additionally
depress the strengths of the line emissions and affect the profile shapes.

\end{appendix}

\end{document}